\documentstyle[11pt]{article}

\newcommand{\sect}[1]{\setcounter{equation}{0}\section{#1}}

\newcommand{\EQ}{\begin{equation}}
\newcommand{\EN}{\end{equation}}
\newcommand{\bea}{\begin{eqnarray}}
\newcommand{\ena}{\end{eqnarray}}

\renewcommand{\a}{\alpha}
\renewcommand{\b}{\beta}

\newcommand{\pa}{\partial}
\newcommand{\g}{\gamma}

\renewcommand{\l}{\lambda}

\newcommand{\s}{\sigma}
\renewcommand{\S}{\Sigma}

\begin{document}

\topmargin 0pt
\oddsidemargin 5mm

\renewcommand{\Im}{{\rm Im}\,}
\newcommand{\NP}[1]{Nucl.\ Phys.\ {\bf #1}}
\newcommand{\PL}[1]{Phys.\ Lett.\ {\bf #1}}
\newcommand{\NC}[1]{Nuovo Cimento {\bf #1}}
\newcommand{\CMP}[1]{Comm.\ Math.\ Phys.\ {\bf #1}}
\newcommand{\PR}[1]{Phys.\ Rev.\ {\bf #1}}
\newcommand{\PRL}[1]{Phys.\ Rev.\ Lett.\ {\bf #1}}
\newcommand{\MPL}[1]{Mod.\ Phys.\ Lett.\ {\bf #1}}
\renewcommand{\thefootnote}{\fnsymbol{footnote}}
\newpage
\begin{titlepage}
\vspace{2cm}
\begin{center}
{\bf{{\Large QUANTUM INTEGRABILITY IN TWO--DIMENSIONAL  }}} \\
{\bf{{\Large  SYSTEMS WITH BOUNDARY}}} \\
\vspace{2cm}
{\large S. Penati and D. Zanon} \\
{\em Dipartimento di Fisica dell' Universit\`{a} di Milano and} \\
{\em INFN, Sezione di Milano, I-20133 Milano, Italy}\\
\end{center}
\vspace{2cm}
\centerline{{\bf{Abstract}}}
\vspace{.5cm}

In this paper we consider affine Toda systems defined on the half--plane
and study the issue of integrability, i.e. the construction of higher--spin
conserved currents in the presence of a boundary perturbation.
First at the classical level we formulate the problem
within a Lax pair approach which allows to determine the general
structure of the boundary perturbation compatible with integrability.
Then we analyze
the situation at the quantum level and compute
corrections to the classical conservation laws in specific examples.
We find that, except for the sinh--Gordon model, the existence of quantum
conserved currents requires a finite renormalization of the boundary potential.

\vfill
\noindent
IFUM--490--FT \hfill \\
\noindent
\hfill {January 1995}
\end{titlepage}
\renewcommand{\thefootnote}{\arabic{footnote}}
\setcounter{footnote}{0}
\newpage

\sect{Introduction}

An integrable field theory possesses higher--spin integrals of motion
which allow to determine exactly the on--shell properties of
the system \cite{zam,b1}. For models defined on the whole two--dimensional
plane the existence of these conserved currents is in one--to--one
correspondence with the existence of a Lax pair formulation
of the equations of motion \cite{b2,b3}. The knowledge of the Lax connection
leads then to a standard, recursive construction of the whole set
of conservation laws.
The analysis is more complicated when the two--dimensional theory
is defined on a manifold with boundary, typically the upper--half
plane. In the presence of the boundary the existence of the conserved
currents in the "bulk" region does not guarantee integrability unless
special boundary conditions are specified appropriately \cite{b4,b5}.

In this paper we consider affine Toda--like theories defined on the half plane,
perturbed by a boundary potential. We study the construction of
classically conserved higher--spin currents
starting from the general setting of a Lax pair approach.
In section 2 we introduce two gauge fields which are defined in the
upper--half plane and contain a non--trivial boundary term. We show that
the compatibility conditions for the corresponding Lax pairs (i.e. curvature of
the gauge fields equal to zero) provide the standard Toda equations of
motion in the bulk region, whereas the boundary conditions on the fields
follow from the requirement that the two curvatures coincide at the
boundary. This condition also fixes the most general structure of the
boundary perturbation compatible with a Lax pair formulation.
Then we discuss how to define a Wilson loop operator on the
whole plane which is time independent and provides an infinite number of
conserved quantities. However, the presence of a boundary does not
guarantee the locality of these currents.
In section 3 we show that
in general the currents obtained from the Lax pair construction consist
in the sum of a bulk term and a boundary term. A {\em local} solution
can be found only if "integrable" boundary conditions are satisfied.

Finally we address the issue of boundary integrability
at the quantum level. In section 4 we explicitly compute the first relevant
exact quantum currents for the boundary sinh--Gordon theory and the $a^{(1)}_2$
Toda model. While in the sinh--Gordon theory quantum integrability is
maintained by a finite renormalization of the currents,
in the $a^{(1)}_2$ case the spin--3 current survives
the quantization only if a nonperturbative renormalization of the
boundary potential is performed.
Section 5 contains our conclusions.

\sect{Lax pair for systems with boundary}

In the upper--half plane we consider a Toda--like system defined
by the euclidean action
\EQ
{\cal S} = \frac{1}{\b^2} \int_{-\infty}^{+\infty} dx_0 \int_0^{+\infty}
dx_1 \left[ \frac12 \pa_{\mu} \vec{\phi} \cdot \pa_{\mu} \vec{\phi}
+ V \right] - \frac{1}{\b^2} \int_{-\infty}^{+\infty} dx_0~ B
\label{1}
\EN
where $B$ is a generic boundary perturbation,
function of the fields but not their derivatives, and $V$ is the affine
Toda potential
\EQ
V = \sum_{j=0}^{N} q_j e^{\vec{\a}_j \cdot \vec{\phi}}
\label{1a}
\EN
The Toda theory under consideration is based
on a Lie algebra ${\cal G}$ of rank $N$, with simple roots $\a_j$,
($j=1,\cdots,N$), $\a_0 = -\sum_{j=1}^N q_j \a_j$, $q_j$ being
the Kac labels ($q_0 = 1$).
The action in (\ref{1}) can be rewritten as an integral on the whole
${\cal R}^2$ plane
\EQ
{\cal S} = \frac{1}{\b^2} \int d^2 x \left\{ \theta(x_1) \left[ \frac12
\pa_{\mu} \vec{\phi} \cdot \pa_{\mu} \vec{\phi} + V \right] -
\delta(x_1)B \right\}
\label{action}
\EN
In standard way one then obtains the Toda equations of motion in
the bulk region
\EQ
\Box \vec{\phi} =  \sum_{j=0}^N q_j \vec{\a}_j e^{\vec{\a}_j \cdot
\vec{\phi}}
\label{2}
\EN
supplemented by the boundary condition
\EQ
\left. \frac{\pa \phi_a}{\pa x_1}\right|_{x_1=0} = - \frac{\pa B}{\pa \phi_a}
\label{3}
\EN

Affine Toda systems without boundary are known to be classically
integrable \cite{b2,b3},
namely there exists an infinite number of spin $\pm n$ currents which
satisfy the conservation laws
\EQ
\bar{\pa} J^{(n)} + \pa \Theta^{(n)} = 0  \qquad \quad
\pa \tilde{J}^{(n)} + \bar{\pa} \tilde{\Theta}^{(n)} = 0
\label{4}
\EN
where we have introduced complex coordinates
\EQ
x = \frac{x_0 + ix_1}{\sqrt{2}} \qquad \quad \bar{x} =
\frac{x_0 - ix_1}{\sqrt{2}}
\EN
and derivatives
\EQ
\pa \equiv \pa_x = \frac{1}{\sqrt{2}} (\pa_0 -i\pa_1) \qquad ~~~
\bar{\pa} \equiv \pa_{\bar x} = \frac{1}{\sqrt{2}} (\pa_0 +i\pa_1) \qquad ~~~
\Box = 2 \pa \bar{\pa}
\EN
The corresponding conserved charges are
\EQ
q^{(n-1)} = \int dx_1 \left( J^{(n)} + \Theta^{(n)} \right)
\qquad \quad \tilde{q}^{(n-1)} =
\int dx_1 \left( \tilde{J}^{(n)} + \tilde{\Theta}^{(n)} \right)
\label{5}
\EN
The integrability of these systems is a consequence of the fact that the
equations of motion (\ref{2}) are the compatibility conditions for a Lax pair
\EQ
(\pa + A) \chi = 0 \qquad \quad (\bar{\pa} + \bar{A} ) \chi = 0
\EN
or equivalently
\EQ
(\pa + \bar{\tilde{A}}) \tilde{\chi} = 0 \qquad \quad (\bar{\pa} + \tilde{A} )
\tilde{\chi} = 0
\EN
where
\bea
&&A(\lambda) = \pa \vec{\phi} \cdot \vec{h} + \lambda \sum_{j=0}^N e_j^+
\qquad \qquad
\tilde{A}(\lambda) = \bar{\pa} \vec{\phi} \cdot \vec{h} + \lambda \sum_{j=0}^N
 e_j^+ \nonumber\\
&&\bar{A}(\lambda) = \bar{\tilde{A}}(\lambda)=
\frac{1}{2\lambda} \sum_{j=0}^N q_j e^{\vec{\a}_j \cdot \vec{\phi}} e_j^-
\label{6}
\ena
Here $\{ \vec{h}, e^+,e^- \}$ are a set of Cartan--Weyl generators
for the Lie algebra ${\cal G}$ and $\lambda$ is the spectral parameter.
The field equations (\ref{2}) are then given by
\EQ
F \equiv [ \pa + A, \bar{\pa} + \bar{A} ] = 0
\EN
or equivalently by
\EQ
\tilde{F} \equiv [ \pa + \bar{\tilde{A}}, \bar{\pa} + \tilde{A} ] = 0
\EN
which are the zero curvature conditions for the ``gauge'' fields $A$, $\bar{A}$
and $\bar{\tilde{A}}$, $\tilde{A}$ respectively.
It follows that by a gauge tranformation
we can always set $A=\bar{A}=0$ on the whole plane so that
the path--ordered Wilson loop
\EQ
W(\lambda) = P~e^{\oint_{\gamma} A}
\EN
does not depend on the time variable. In the same way we
can define a corresponding Wilson
loop in terms of $\bar{\tilde{A}}$, $\tilde{A}$.
Finally expanding $W(\lambda)$ and $\tilde{W}(\lambda)$ as a series in
$\lambda$ one obtains the infinite set of conserved quantities in (\ref{4}).

If the theory is defined on the semi--infinite plane $x_1 \ge 0$, it is easy
to show that from the local conservation laws (\ref{4}) valid now in the
upper--half plane one can still define conserved charges if the
following boundary conditions are satisfied
\EQ
\left. J_1^{(n)} \right|_{x_1=0} \equiv
\left. i\left( J^{(n)} - \tilde{J}^{(n)} - \Theta^{(n)} + \tilde{\Theta}^{(n)}
\right) \right|_{x_1=0} = -\pa_0 \Sigma_0^{(n)}
\label{9}
\EN
with $\Sigma_0$ any local function of the
the fields at $x_1=0$.
The corresponding conserved charge is given by
\EQ
q^{(n-1)} = \int_0^{+\infty} dx_1 J_0^{(n)} ~~+~ \Sigma_0^{(n)}
\label{8}
\EN
where $J_0^{(n)} = J^{(n)} + \tilde{J}^{(n)} + \Theta^{(n)} +
\tilde{\Theta}^{(n)}$. It has been
shown that the condition (\ref{9}) restricts the class
of boundary perturbations $B$ \cite{b4,b5}. In the sine--Gordon case
the classical integrability of the system is maintained for
$B = \gamma \cos{\frac{\phi - \phi_0}{2}}$ where $\g$ and
$\phi_0$ are arbitrary constants, whereas in the $a_n^{(1)}$, $n>1$ case
the integrable perturbation is $B = \sum_{j=0}^N d_j e^{\frac12 \vec{\a}_j
\cdot \vec{\phi}}$ where the coefficients $d_j$ must satisfy $d_j^2 = 4$.

Our aim now is to recast the above results in a Lax pair framework
suitable for the description of
two--dimensional systems in the presence of a boundary. We consider
a theory described by the action in (\ref{action}) with $V$ given in
(\ref{1a}) and a generic boundary perturbation of the form
$B = \sum_{j=0}^N f_j$ where $f_j$ is a function of the $j$--th root
of the algebra.
The underlying idea is to introduce appropriate gauge connections defined
in the bulk and at the boundary,
such that in the interior and at the border the field equations
in (\ref{2}) and (\ref{3})
correspond to a zero curvature condition of the gauge fields.
To this end we start again from a ``chiral''
curvature F  and its ``antichiral'' counterpart $\tilde{F}$ defined
in the upper--half plane and at the border as
\bea
&& F = \pa \bar{A} - \bar{\pa} A + [A,\bar{A}] \nonumber \\
&& \tilde{F} = \pa \tilde{A} - \bar{\pa} \bar{\tilde{A}} - [\tilde{A},
\bar{\tilde{A}}]
\label{10}
\ena
where now the gauge fields are chosen of the form
\bea
&& A = \theta(x_1) \left[ \pa \vec{\phi} \cdot \vec{h} + \lambda
\sum_{j=0}^N e_j^+ \right] \qquad \quad
\tilde{A} = \theta(x_1) \left[ \bar{\pa} \vec{\phi} \cdot \vec{h}
+ \lambda \sum_{j=0}^N e_j^+ \right] \nonumber \\
&& \bar{A} = \bar{\tilde{A}} = \frac{1}{2\lambda} \theta(x_1)~ \sum_{j=0}^N
q_j e^{\vec{\a}_j \cdot \vec{\phi}}  ~e_j^-  - \frac{1}{2\lambda} \delta(x_1)~
\sum_{j=0}^N f_j ~e_j^-
\label{11}
\ena
With this choice (cfr. eq. (\ref{6}))
the correct Toda bulk equations of motion (\ref{2}) are given by
$F=0$, $\tilde{F}=0$ in the interior region.
At the boundary $x_1=0$, we impose the condition $F=\tilde{F}$
so that
the curvature $\tilde{F}$ can be interpreted as the analytic continuation of
$F$ in the lower--half plane.
By computing explicitly $\left. (F-\tilde{F})\right|_{x_1=0}$
we obtain
\bea
\left. (F-\tilde{F})\right|_{x_1=0} &=&  \left[
-i\frac{(\pa - \bar{\pa})}{\sqrt{2}}\vec{\phi} - \frac12 \sum_{j=0}^N
f_j~\vec{\a}_j \right] \cdot \vec{h} \nonumber \\
&~&~~- \frac{1}{2\lambda} \sum_{j=0}^N \left[ \frac{\pa f_j}{\pa
\vec{\phi}} -\frac12 f_j~ \vec{\a}_j \right]
\cdot (\pa + \bar{\pa}) \vec{\phi}~ e_j^-
\label{12}
\ena
where we have used the relations $\pa \theta(x_1) =
-\bar{\pa}\theta(x_1) = -\frac{i}{\sqrt{2}} \delta(x_1)$ and
$\theta(x_1) \delta(x_1) = \frac12 \delta(x_1)$. Therefore
$\left. (F-\tilde{F})\right|_{x_1=0}=0$
if
\EQ
\frac{\pa f_j}{\pa \vec{\phi}} = \frac12 f_j \vec{\a}_j
\label{13}
\EN
and
\EQ
\left. \frac{ \pa \vec{\phi}}{\pa x_1} \right|_{x_1=0} = - \frac12
\sum_{j=0}^N  f_j~ \vec{\a}_j
\label{14}
\EN
The condition in eq. (\ref{13}) implies that the boundary
perturbation must be of the form
\EQ
B = \sum_{j=0}^N d_j e^{\frac12 \vec{\a}_j \cdot \vec{\phi}}
\label{13a}
\EN
for arbitrary coefficients $d_j$,
and the relation (\ref{14}) gives the boundary equation (\ref{3}).
We emphasize
that this result holds for all Toda systems and generalizes what
obtained in Ref. \cite{b5}. At this
stage no further restriction needs be imposed on the boundary perturbation
$B$.

The bulk zero--curvature conditions $F=0$, $\tilde{F}=0$ and the requirement
$F=\tilde{F}$ at the boundary allow now to proceed and construct
Wilson loop operators.
Thus we define
\bea
W(\lambda) &=& P~e^{\oint_C A} = e^{\int_D F} \nonumber \\
\tilde{W}(\lambda) &=& P~e^{\oint_C \tilde{A}} = e^{\int_D \tilde{F}}
\label{15}
\ena
where $C$ is any close contour in the upper--half plane enclosing the
region $D$. Due to the bulk
conditions $F=0$, $\tilde{F}=0$ these Wilson operators are equal to 1
in the upper--half plane and do not depend on
the choice of $C$. Moreover since $\tilde{F}$ is by definition the analytic
continuation of $F$ in the lower--half plane, $\tilde{W}$ can be rewritten
as
\EQ
\tilde{W}(\lambda) = e^{\int_{\bar{D}} F}
\EN
where $\bar{D}$ is a region enclosed by a contour $\bar{C}$ in the lower
half plane. Therefore the operator $W(\lambda)\tilde{W}
(\lambda)$, with $D$ and $\bar{D}$ as shown in Fig. 1,
can be defined on the whole circle and it is there
equal to 1 except for possible boundary
contributions. In fact the condition $\left. (F -\tilde{F}) \right|_{x_1=0}
=0$ guarantees that boundary effects are not present and we are left
with a Wilson operator which is well defined on the whole plane and does
not depend on time. We can then expand it in a power
series in $\lambda$: if the coefficients are local in
the fields they provide the conserved charges of the theory.
As we will show in the next section, in the presence of a boundary
the requirement of
locality is not automatically satisfied and it may impose additional
restrictions on the boundary
perturbation $B$.

\sect{Classical conserved currents}

Now we want to show that the procedure outlined in the previous section
provides a consistent way to determine conserved
currents for systems with boundary.
In general, given the gauge connections as in eq. (\ref{11}),
we will find quantities that consist in the sum of a bulk
and a boundary contribution
\bea
&& J' = \theta(x_1) J + \delta(x_1) J_B \qquad \quad
\Theta' = \theta(x_1) \Theta +\delta(x_1) \Theta_B \nonumber \\
&& \tilde{J}' = \theta(x_1) \tilde{J} + \delta(x_1) \tilde{J}_B
\qquad \quad \tilde{\Theta}' = \theta(x_1) \tilde{\Theta} +
\delta(x_1) \tilde{\Theta}_B
\label{33}
\ena
Here $J$, $\Theta$ and $\tilde{J}$, $\tilde{\Theta}$ are the
currents which satisfy standard conservation laws in the bulk region
( cfr. eq.(\ref{4}) )
\EQ
\pa_0 (J +\Theta)+ i\pa_1 (J-\Theta)=0 \qquad \qquad
\pa_0 (\tilde{J}+\tilde{\Theta}) -i\pa_1(\tilde{J}-\tilde{\Theta})=0
\label{34}
\EN
whereas $J_B$, $\Theta_B$, $\tilde{J}_B$ and $\tilde{\Theta}_B$ are the
boundary terms.
The Lax pair approach leads to generalized currents defined on the
whole plane, with components
\bea
&& J_0' \equiv J'+ \tilde{J}' +\Theta' +\tilde{\Theta}'
=  \theta(x_1) J_0 + \delta(x_1) \Sigma_0 \nonumber \\
&& J_1' \equiv i(J' -\tilde{J}'-\Theta' +\tilde{\Theta}')
= \theta(x_1) J_1 + \delta(x_1) \Sigma_1
\label{36}
\ena
where we have defined
\bea
&&J_0 \equiv J+ \tilde{J} +\Theta +\tilde{\Theta}
\qquad \qquad \quad
\Sigma_0 \equiv J_B + \tilde{J}_B + \Theta_B + \tilde{\Theta}_B
\nonumber\\
&& J_1 \equiv i(J -\tilde{J}-\Theta +\tilde{\Theta})
\qquad \qquad \S_1 \equiv i(J_B - \tilde{J}_B  -\Theta_B + \tilde{\Theta}_B)
\ena
It is easy to check that the conservation law
\EQ
\pa_0 J_0' + \pa_1 J_1' =0
\label{conserv}
\EN
which can be rewritten as
\EQ
\pa_0 (J'+\Theta') +i\pa_1 (J'-\Theta') = -\pa_0 (\tilde{J}' +
\tilde{\Theta}') +i\pa_1 (\tilde{J}' - \tilde{\Theta}')
\label{37a}
\EN
holds in the bulk region whenever (\ref{34}) are valid, whereas at the
boundary it gives
\EQ
J_1 |_{x_1 =0}=\lim_{x_1 \to 0}
{}~i(J-\tilde{J} - \Theta + \tilde{\Theta}) \equiv -\pa_0 \S_0
\label{35}
\EN
Thus it is clear that one can construct local boundary terms in the currents
only if the boundary condition in eq. (\ref{3}) allows to express
$J_1 |_{x_1 =0}$ as a time derivative of a functional
of the fields. If this is the case, from the conservation equation
in (\ref{conserv}) one obtains the corresponding charge
\EQ
q= \int_{-\infty}^{+\infty} dx_1~J_0'
\label{37}
\EN
which in fact coincides with the one defined in (\ref{8}).

We consider now as a specific example the
sinh--Gordon theory defined on the half plane. It is the simplest
affine Toda system which contains
a single scalar field. The corresponding action is obtained from
eq. (\ref{1}), setting $N=1$, $\a_1=1$, $\a_0=-1$, and $q_0=q_1=1$
in (\ref{1a}) and (\ref{13a}) so that
\EQ
V= e^{-\phi} +e^{\phi}\qquad \qquad
B= d_0 e^{-\frac{\phi}{2}} +
d_1e^{\frac{\phi}{2}}
\label{pot}
\EN
For the gauge connections in equation (\ref{11})
we choose the following
realization of the $SU(2)$ algebra: $h= \frac12 \s_x$,
$e_0^- = e_1^+ = -\frac{1}{2\sqrt{2}} (\s_z -i\s_y)$ and
$e_0^+ = e_1^- =-\frac{1}{2\sqrt{2}} (\s_z +i\s_y )$ where $\{\s_x, \s_y,
\s_z\}$ are the Pauli matrices.

In the interior region $x_1>0$ the determination of the currents proceeds
in standard manner. One writes the Lax equations
\bea
&& (\pa + A)\chi =0 \qquad \qquad (\bar{\pa} + \bar{A} )\chi =0 \nonumber \\
&& (\pa + \bar{\tilde{A}}) \tilde{\chi} =0 \qquad \qquad (\bar{\pa} +
\tilde{A} )\tilde{\chi} = 0
\label{16}
\ena
where $\chi= (\chi_1, \chi_2)$ and $\tilde{\chi}=(\tilde{\chi}_1,
\tilde{\chi}_2)$ are two components vectors. Then one introduces
new variables $V=\chi_2/\chi_1$, $U =\log{\chi_1}$, so that (\ref{16})
give
\bea
&& \pa V + \sqrt{2} \l \theta(x_1) V - \frac12 \theta(x_1)\pa \phi V^2
+\frac12 \theta(x_1) \pa \phi =0 \nonumber \\
&& \bar{\pa}V +
\frac{1}{\sqrt{2}\l} \theta(x_1)\left[ \cosh{\phi}~ V
+\frac12 \sinh{\phi}(V^2 +1) \right] =0
\label{17}
\ena
and
\bea
&& \pa U +\frac12 \theta(x_1) \pa \phi~ V -\frac{\l}{\sqrt{2}} \theta(x_1)
=0 \nonumber \\
&& \bar{\pa} U -\frac{1}{2\sqrt{2}\l} \theta(x_1) \left[ \sinh{\phi}~ V +
\cosh{\phi} \right] =0
\label{18}
\ena
and analogous equations for $\tilde{V}= \tilde{\chi}_2/\tilde{\chi}_1$
and $\tilde{U}= \log{\tilde{\chi}_1}$ exchanging $\pa$ with $\bar{\pa}$
in (\ref{17}), (\ref{18}).

Thus, as a consequence of the two conditions
$F\chi =0$ and $\tilde{F}\tilde{\chi}=0$ valid away from
the boundary, from (\ref{18}) and the corresponding ones
for $\tilde{U}$, $\tilde{V}$ , using
$ \pa \bar{\pa} U = \bar{\pa} \pa U$
and $ \pa \bar{\pa}\tilde{ U} = \bar{\pa} \pa \tilde{U}$
one obtains two conservation equations in the bulk region (cfr. (\ref{34}))
\bea
\bar{\pa} \left( \frac12 \theta(x_1) \pa \phi~ V -\frac{\l}{\sqrt{2}}
\theta(x_1) \right) = -\pa \left( \frac{1}{2\sqrt{2}\l} \theta(x_1) [
\sinh{\phi} ~V + \cosh{\phi} ] \right) \nonumber \\
\pa \left( \frac12 \theta(x_1) \bar{\pa}\phi~ \tilde{V}-\frac{\l}{\sqrt{2}}
\theta(x_1) \right) = -\bar{\pa} \left( \frac{1}{2\sqrt{2}\l} \theta(x_1)
[\sinh{\phi}~ \tilde{V} + \cosh{\phi}] \right)
\label{19}
\ena
As in the case of Toda systems without boundary one then expands
\EQ
 V = \sum_{n=1}^{\infty} \frac{a_n}{(\sqrt{2}\l)^n} \qquad
\qquad \tilde{V} = \sum_{n=1}^{\infty} \frac{\tilde{a}_n}{(\sqrt{2}\l)^n}
\label{23}
\EN
and from (\ref{19}) an
infinite number of conservation laws is generated. The coefficients
$a_n$ are determined recursively from the first equation in (\ref{17})
\bea
&&a_1= -\frac{1}{2} \pa \phi \nonumber\\
&&a_n = -\pa a_{n-1} + \frac12 \pa \phi \sum_{j=1}^{n-1} a_j a_{n-1-j}
{}~~~~~~,~~~~~~n>1
\label{29}
\ena
with similar expressions for  $\tilde{a}_n$.

At the boundary $x_1=0$ we impose the additional
condition $F\chi = \tilde{F} \tilde{\chi}$. It is easy to show that
this amounts to the following equations
\bea
&& \bar{\pa} \left[ \sqrt{2} \l \theta(x_1) V -\frac12 \theta(x_1) \pa \phi~
V^2
+\frac12 \theta(x_1) \pa \phi \right] + \pa \left[ \sqrt{2} \l \theta(x_1)
\tilde{V}-\frac12 \theta(x_1) \bar{\pa} \phi~ \tilde{V}^2 + \frac12 \theta(x_1)
\bar{\pa} \phi \right] \nonumber \\
&& -\pa \left[ \frac{1}{\sqrt{2}\l} \theta(x_1) ( \cosh{\phi} ~V + \frac12
\sinh{\phi}~(V^2+1) ) \right]  \nonumber \\
&& -\bar{\pa} \left[ \frac{1}{\sqrt{2}\l}
\theta(x_1) (\cosh{\phi}~\tilde{V} + \frac12 \sinh{\phi} ~(\tilde{V}^2+1)
\right] \nonumber \\
&& - \frac{1}{4\l} \pa_0 \left\{ \delta(x_1) \left[ (d_0 e^{-\frac{\phi}{2}}
+d_1 e^{\frac{\phi}{2}} )(V+\tilde{V}) + \frac12 (d_1 e^{\frac{\phi}{2}}
- d_0 e^{-\frac{\phi}{2}} )(V^2 +\tilde{V}^2 +2) \right] \right\}
=0
\label{21}
\ena
and
\bea
&& \bar{\pa} \left( \frac12 \theta(x_1) \pa \phi~ V \right) + \pa
\left( \frac12 \theta(x_1) \bar{\pa} \phi ~\tilde{V} \right) \nonumber \\
&& + \pa \left[ \frac{1}{2\sqrt{2}\l} \theta(x_1) (\sinh{\phi}~ V +
\cosh{\phi})
\right]
+ \bar{\pa} \left[ \frac{1}{2\sqrt{2}\l} \theta(x_1) (\sinh{\phi}~
\tilde{V} + \cosh{\phi} ) \right] \nonumber \\
&& + \frac{1}{4\l} \pa_0 \left\{ \delta(x_1) \left[ \frac12 (d_1
e^{\frac{\phi}{2}} - d_0 e^{-\frac{\phi}{2}})(V+\tilde{V}) + (d_0
e^{-\frac{\phi}{2}} + d_1 e^{\frac{\phi}{2}}) \right] \right\} =0
\label{22}
\ena
which are obviously well defined on the whole plane. In particular
we note that (\ref{22}) expresses the conservation equation in the same
form as in (\ref{37a}).
Now in order to account for the boundary contributions correctly,
in the expansion (\ref{23}) we have to replace
$a_n \to \theta(x_1)a_n + \delta(x_1) b_n$ and similarly for
$\tilde{a}_n$, with the bulk coefficients given in (\ref{29})
and the boundary terms $b_n$, $\tilde{b}_n$ to be determined
by the equations (\ref{21}), (\ref{22}).

For example one can easily  show that
for the stress--energy tensor and the spin--3
current ( which is a total derivative ), the boundary conditions (\ref{21})
and (\ref{22}) are automatically satisfied by $b_1=\tilde{b}_1=0$ and
$b_2=\tilde{b}_2=0$. The first nontrivial case is
at spin--4. The bulk contributions $a_3$, $\tilde{a}_3$
are determined using (\ref{29}). At the boundary, the equations
(\ref{21}), (\ref{22}) to second and third order in $\frac{1}{\l}$
respectively, give
\bea
&& \left. \pa_0 (b_3 +\tilde{b}_3) = -\frac{1}{2} \pa_0 \left[ (\pa \phi +
\bar{\pa} \phi) (d_0 e^{-\frac{\phi}{2}}
+ d_1 e^{\frac{\phi}{2}}) \right] \right|_{x_1 =0}
\nonumber \\
&~&~~~~~~~~~~~~~~~ \left. +i \left\{ (\pa^3 \phi - \bar{\pa}^3 \phi) +
\cosh{\phi} (\pa \phi - \bar{\pa} \phi) \right\} \right|_{x_1 =0}
\label{30}
\ena
and
\bea
&& \left. \pa_0 (\pa \phi b_3 + \bar{\pa} \phi \tilde{b}_3) = -\frac12
\pa_0 \left[ (\pa^2 \phi + \bar{\pa}^2 \phi) (d_1 e^{\frac{\phi}{2}} -
d_0 e^{-\frac{\phi}{2}}) \right] \right|_{x_1 =0} \nonumber \\
&& \left. -i\left \{\frac{1}{4}( (\pa \phi)^4 - (\bar{\pa} \phi)^4) -
(\pa \phi \pa^3 \phi - \bar{\pa} \phi \bar{\pa}^3 \phi)
+ \sinh{\phi}~ (\pa^2 \phi - \bar{\pa}^2 \phi) \right\} \right|_{x_1 =0}
\label{31}
\ena
These equations define a spin--4 conserved current with the correct
behavior at the boundary if they admit a {\em local} solution, i.e. if
the quantities in curly brackets are time derivatives of
local expressions.
While in equation (\ref{30}) this is always true being the curly
brackets equal to the $1/\l^2$ coefficient of $\pa_0( V +\tilde{V})$,
the expression in curly brackets in equation (\ref{31}) is
exactly the bulk component $J_1$ of the spin--4 current evaluated at the
boundary.
Thus the locality condition  requires that $\left. J_1
\right|_{x_1=0}$ be the time derivative of a given function of the fields
evaluated at the boundary.
It is easy to show that the same pattern repeats itself at any spin level,
namely for the spin--$n$ current the boundary conditions have the general
form
\bea
&& \pa_0 (b_{n-1} +\tilde{b}_{n-1}) = \pa_0 X \nonumber \\
&& \pa_0 (\pa \phi ~b_{n-1} + \bar{\pa} \phi ~\tilde{b}_{n-1})
= \left. -J_1^{(n)}
\right|_{x_1=0} + \pa_0 Y
\label{32}
\ena
with $X$ and $Y$ local functional of the fields.
Notably in the sinh--Gordon case this condition is automatically satisfied by
the $B$ potential in (\ref{pot}).

The previous arguments can be generalized to the $a_n^{(1)}$ Toda systems.
We have checked explicitly for the spin--3 current of the $n=2$ case that
locality of the $b$ and $\tilde{b}$ coefficients requires once again
$\left. J_1\right|_{x_1=0}$ to be a time derivative. As
shown in Ref. \cite{b5} this condition is not automatically satisfied by
a perturbation of the form (\ref{13a}) but one has to choose $d_j^2 = 4$,
$j=0,1,2$.

\sect{Quantum conserved currents}

In this section we investigate the problem of quantum integrability of
systems with boundary by studying the renormalization of the
classical conservation laws.
As described above in the case of systems defined on the half plane
classical conserved currents give rise to physical conserved charges if the
extra condition (\ref{9}) is satisfied at the boundary.
However at the quantum level the conservation laws (\ref{4}) and the boundary
condition (\ref{9}) might be affected by anomalies. We compute these potential
anomalous terms and show that they can be reabsorbed in a quantum
redefinition of the currents. In addition
a finite renormalization of the
boundary potential is in general necessary in order to maintain integrability.

The calculation is most easily performed using massless perturbation theory.
For the action in (\ref{1}) the massless propagator is defined by
the equations
\EQ
\frac{1}{\b^2} \Box G_{ij}(x,x') = - \delta_{ij} \delta^{(2)}(x-x')
\qquad \quad
\left. \frac{\pa G_{ij}}{\pa x_1}(x,x') \right|_{x_1=0} = 0
\label{40}
\EN
Thus using the relation $\bar{\pa}(1/x)  = 2\pi \delta^{(2)}(x)$ one finds
\EQ
G_{ij}(x,x') = -\frac{\b^2}{4\pi} \delta_{ij}
\left[ \log{2|x-x'|^2} + \log{2|x-\bar{x}'|}^2 \right]
\label{41}
\EN
Then one treats ${\cal S}_{int} =  \frac{1}{\b^2}
\int_{-\infty}^{+\infty} dx_0 \int_0^{+\infty} dx_1 V$, with $V$ the
affine Toda potential (\ref{1a}) and ${\cal S}^B_{int} = -\frac{1}{\b^2}
\int_{-\infty}^{+\infty} dx_0 B$, with $B$ the boundary perturbation,
as interaction terms. In $V$ and $B$ the exponentials are normal ordered
so that perturbative calculations are free from ultraviolet divergences.

At the quantum level the conservation equations become
\EQ
\bar{\pa} \left\langle J^{(n)}(x,\bar{x}) \right\rangle \equiv \bar{\pa}
\left\langle J^{(n)}(x,\bar{x}) ~e^{- {\cal S}_{int}} \right\rangle_0 =
\pa \left\langle \Theta^{(n)} \right\rangle
\label{38}
\EN
for $x_1>0$ and
\EQ
\left. \left\langle J^{(n)}_1(x,\bar{x}) \right\rangle \right|_{x_1=0}
\equiv \left. \left\langle J^{(n)}_1(x,\bar{x}) ~e^{- {\cal S}_{int}^B}
\right\rangle_0 \right|_{x_1=0} = ~\pa_0-{\rm derivative}
\label{39}
\EN
at $x_1=0$.
Anomalous contributions would correspond to {\em local} terms
obtained by Wick contracting the currents with the exponentials
in (\ref{38}) and (\ref{39}). Mixing between bulk and boundary
interactions have not been included since they would always
produce non--local expressions.

First we consider $\bar{\pa} \left\langle J^{(n)} \right\rangle $
in (\ref{38}). Being interested only in local contributions it is
sufficient to expand the exponential to first order in $S_{int}$.
Indeed performing the Wick contractions with
the propagator (\ref{41}), we obtain terms of the form
\EQ
\bar{\pa}_x~ \int d^2 w~ ~{\cal M}(x,\bar{x}) \left[
\frac{1}{(x-w)^k} + \frac{1}{(x-\bar{w})^k} \right] ~{\cal N}(w,\bar{w})
\label{42}
\EN
where ${\cal M}$, ${\cal N}$ are products of the fields and their
$\pa$--derivatives and the integration is performed in the upper--half plane.
Now local expressions arise using in the half plane the relation
\EQ
\bar{\pa}_x \frac{1}{(x-w)^k} = \frac{2\pi}{(k-1)!} \pa_w^{k-1} \delta^{(2)}
(x-w)
\label{43}
\EN
The contributions obtained in this way either can be rewritten as total
$\pa$--derivatives and then give corrections to the quantum
trace in (\ref{38}), or must be reabsorbed in a renormalization of the
classical current $J^{(n)}$.
Following this procedure, which is exact to all--loop orders, one determines
the quantum current $J^{(n)}$ and its corresponding trace
$\Theta^{(n)}$ defined for $x_1>0$.

Then one has to study the condition (\ref{39}) at the boundary,
using the exact quantum expression of $J_1^{(n)}=i(J^{(n)}-\tilde{J}^{(n)}-
\Theta^{(n)}+\tilde{\Theta}^{(n)})$.
Again we are looking for potential anomalies, namely for local terms
which are not expressible as total $\pa_0$--derivatives.
In this case one needs consider Wick contractions with higher--order
terms in the expansion of the boundary potential. Tipically, expanding
the exponential in (\ref{39}) to first order, we produce terms with
the following general structure
\EQ
\lim_{x_1 \to 0}
\int_{-\infty}^{+\infty} dw_0 ~ \left[{\cal P}(x,\bar{x})
 \left( \frac{1}{(x-w)^k}
+ \frac{1}{(x-\bar{w})^k} \right) -\tilde{{\cal P}}(x,\bar{x})
\left( \frac{1}{(\bar{x}-\bar{w})^k} + \frac{1}{(\bar{x}-w)^k} \right)
 \right]{\cal Q}(w_0)
\label{44}
\EN
Here
${\cal P}$ is a function of $\pa^p \phi$ and $\tilde{{\cal P}}$
correspondingly of $\bar{\pa}^p \phi$. Since $w=\bar{w}$ ($w_1=0$)
(\ref{44}) gives
\EQ
2(\sqrt{2})^k
{}~\lim_{x_1 \to 0} \int_{-\infty}^{+\infty} dw_0 ~ \left[{\cal P}(x,\bar{x})
\frac{1}{(x_0-w_0 +ix_1)^k}  -\tilde{{\cal P}}(x,\bar{x})
\frac{1}{(x_0-w_0-ix_1)^k} \right]{\cal Q}(w_0)
\label{45}
\EN
Then in order to identify local boundary contributions
we isolate in ${\cal P}$ and $\tilde{{\cal P}}$ terms which are identical
and use the relation
\EQ
\lim_{x_1 \to 0^+} \left( \frac{1}{(x_0-w_0 -ix_1)^k} -
\frac{1}{(x_0-w_0+ix_1)^k} \right) = \frac{2 \pi i}{(k-1)!} \pa_{w_0}^{k-1}
\delta(x_0 -w_0)
\label{47}
\EN
In a similar manner contractions with higher--order factors in the expansion
of the boundary interaction give rise to local terms whenever
the number of delta functions produced in the limit $x_1 \to 0$
equals the number of integrations.

At this stage one has to analyze the local contributions which cannot
be written as $\pa_0$--derivatives
of suitable expressions, and understand whether they
correspond to real boundary anomalies or if
it is possible to eliminate them by coupling--constant dependent
modifications of the boundary potential.
It is easy to show that for any Toda theory defined on
the half plane the
quantum stress--energy tensor satisfies
the condition (\ref{9}) without any new condition on the boundary
perturbation $B$.
In general restrictions arise when the construction of
quantum higher--spin conservation laws is attempted.
We have performed the explicit calculation for two cases, namely the
spin--4 current of the sinh--Gordon system and the spin--3 current of the
$a_2^{(1)}$ Toda model. We illustrate our results in the next two subsections.

\subsection{The sinh--Gordon model}

The first nontrivial classical conserved current for the sinh--Gordon model
with boundary is the spin--4 current. In order to determine its quantum version
we consider the general expression
\EQ
J^{(4)} = \frac{A}{4} (\pa \phi)^4 + \frac{D}{2} (\pa^2 \phi)^2
\label{48}
\EN
and study $\bar{\pa} \left\langle J^{(4)} \right\rangle$ following the
procedure outlined above. Thus we evaluate
\bea
&& \bar{\pa} \left\langle (\pa^2 \phi)^2 \right\rangle \sim
-\frac{1}{\b^2}  \int d^2 w ~\left\{ 2 ~\pa^2\phi
{}~\frac{\b^2}{4\pi} \left[ \bar{\pa} \frac{1}{(x-w)^2} + \bar{\pa}
\frac{1}{(x-\bar{w})^2} \right] \frac{\pa V}{\pa \phi} ~\right. \nonumber \\
&~&~~~~~~~~~~~~~~~~~~~\left. +
\left( \frac{\b^2}{4\pi} \right)^2 \left[ \bar{\pa} \frac{1}{(x-w)^4} +
\bar{\pa} \frac{1}{(x-\bar{w})^4} \right]\frac{\pa^2 V}{\pa \phi^2} \right\}
\label{49}
\ena
where the integrations are performed in the upper--half plane.
Using the relation (\ref{43}) we obtain
\bea
\bar{\pa} \left\langle (\pa^2 \phi)^2 \right\rangle &\sim&
-\int d^2w ~\pa^2\phi \left[ \pa \delta^{(2)}(x-w) +
\pa \delta^{(2)}(x-\bar{w}) \right]\frac{\pa V}{\pa \phi} ~ \nonumber \\
&~&- \frac{\a}{24}\int d^2w~ \left[ \pa^3
\delta^{(2)}(x-w) + \pa^3 \delta^{(2)}(x-\bar{w}) \right]
\frac{\pa^2 V}{\pa \phi^2}
\label{50}
\ena
having defined $\a \equiv \frac{\b^2}{2\pi}$.
In the upper--half plane only
one of the two delta functions contributes and we find
\bea
&& \bar{\pa} \left\langle (\pa^2 \phi)^2 \right\rangle \sim
\pa (\frac{\pa V}{\pa \phi}) \pa^2 \phi + \frac{\a}{24} \pa^3
\frac{\pa^2 V}{\pa \phi^2} \nonumber \\
&& = -\frac12 \frac{\pa^3 V}{\pa \phi^3} (\pa \phi)^3 +\frac12 \pa\left(
\frac{\pa^2 V}{\pa \phi^2} (\pa \phi)^2 \right) +\frac{\a}{24} \pa^3
\frac{\pa^2 V}{\pa\phi^2}
\label{51}
\ena
Computing in the same manner the term $\bar{\pa} \left\langle (\pa \phi)^4
\right\rangle$ the total result is
\bea
&& \bar{\pa} \left\langle J^{(4)} \right\rangle = \nonumber \\
&& = \pa \left[ \frac{D}{4} \frac{\pa^2 V}{\pa \phi^2} (\pa \phi)^2 +
\a ~\frac{D}{48}~ \pa^2 \frac{\pa^2 V}{\pa \phi^2} +
\a^2~\frac{A}{32}~  \frac{\pa^4 V}{\pa \phi^4} (\pa \phi)^2 +
\a^3~\frac{A}{384}~
\pa^2 \frac{\pa^4 V}{\pa \phi^4} \right] \nonumber \\
&& +\frac12\left[ A \frac{\pa V}{\pa \phi} +\left( - \frac{D}{2} +
\a~\frac{3}{4} A~  \right) \frac{\pa^3 V}{\pa \phi^3} + \a^2~\frac{A}{16}~
\frac{\pa^5 V}{\pa \phi^5} \right] (\pa \phi)^3
\label{52}
\ena
Therefore in the bulk region the conservation law is not anomalous if the
coefficients $A$ and $D$ satisfy
\EQ
A \frac{\pa V}{\pa \phi} +\left( - \frac{D}{2} +\a~
\frac{3A}{4}  \right) \frac{\pa^3 V}{\pa \phi^3} +\a^2~ \frac{A}{16}~
\frac{\pa^5 V}{\pa \phi^5} = 0
\label{53}
\EN
Using now the explicit expression for the sinh--Gordon potential $V =2
\cosh{\phi}$ we obtain
\EQ
D = 2A \left( 1+\frac{3}{4} \a + \frac{\a^2}{16} \right)
\label{54}
\EN
We note that in the classical limit ($\a =0$) this coincides with the
standard relation \cite{b6}.
The corresponding quantum trace is given in (\ref{52}) and can be rewritten as
\EQ
\Theta^{(4)} = -\frac{1}{4} \left( D + \a \frac{D}{12} + \a^2 \frac{A}{8} +
\a^3 \frac{A}{96} \right) \frac{\pa^2 V}{\pa \phi^2} ~(\pa \phi)^2 -
\frac{\a}{48} \left( D+\a^2 \frac{A}{8} \right) \frac{\pa V}{\pa \phi}
{}~\pa^2 \phi
\label{56}
\EN
where the coefficients $A$ and $D$ satisfy (\ref{54}).
The currents with opposite spin $\tilde{J}^{(4)}$,
$\tilde{\Theta}^{(4)}$ have similar expressions with $\pa$--derivatives
substituted by $\bar{\pa}$--ones.

Now we consider the boundary condition in equation ({\ref{39}) and
evaluate
\EQ
\left. \left\langle i(J^{(4)} - \tilde{J}^{(4)} -
\Theta^{(4)} +\tilde{\Theta}^{(4)})~e^{\frac{1}{\b^2} \int dw_0 B}
\right\rangle\right|_{x_1=0}
\EN
Local contributions from $\left\langle\Theta^{(4)} -
\tilde{\Theta}^{(4)} \right\rangle$
arise only from the first order expansion in $B$. First we compute
\bea
&& \left.
\left\langle \frac{\pa^2 V}{\pa \phi^2} (\pa \phi)^2 - \frac{\pa^2 V}{\pa
\phi^2} (\bar{\pa} \phi)^2 \right\rangle \right|_{x_1=0}  \nonumber \\
&& \sim \lim_{x_1 \to 0}
\frac{1}{\b^2} \int dw_0~ 4 ~
\frac{\pa B}{\pa \phi}~ \frac{\pa^2 V}{\pa
\phi^2} \left( -\frac{\b^2}{4\pi} \right) \left[
\frac{(\pa_0 + i\pa_1) \phi}{x_0-w_0+ix_1}
- \frac{(\pa_0-i\pa_1) \phi}{x_0-w_0-ix_1} \right] \nonumber \\
&& \sim 2i~ \frac{\pa B}{\pa \phi} ~\frac{\pa^2 V}{\pa \phi^2}~ \pa_0 \phi
\label{57}
\ena
where the local contribution arises from the terms proportional to
$\pa_0 \phi$ by using the relation (\ref{47}).
Performing an analogous calculation for the other term in the trace
finally we obtain
\bea
&& \left\langle \Theta^{(4)} - \tilde{\Theta}^{(4)} \right\rangle \\
&& = -\frac{i}{2} \left( D + \a \frac{D}{12} +\a^2 \frac{A}{8} + \a^3
\frac{A}{96} \right) \frac{\pa^2
V}{\pa \phi^2}~ \frac{\pa B}{\pa \phi} ~\pa_0 \phi -i \frac{\a}{24} \left(
D + \a^2 \frac{A}{8} \right)
\frac{\pa V}{\pa \phi}~ \frac{\pa^2 B}{\pa \phi^2}~ \pa_0 \phi
\nonumber
\label{58}
\ena
The calculation of $\left\langle J^{(4)} - \tilde{J}^{(4)}\right\rangle$
is more complicated since in this case local terms
arise up to third order in the $B$ expansion. As an example of higher--order
contributions we describe here the computation of
$\left\langle (\pa \phi)^4 - (\bar{\pa} \phi)^4
\right\rangle$.  To first order in $B$ the Wick contractions,
up to total $\pa_0$--derivatives, give
\bea
&& \lim_{x_1 \to 0} \frac{1}{\b^2} \int dw_0
\left\{ 8 \sqrt{2} \frac{\pa B}{\pa \phi} \left( -\frac{\b^2}{4\pi} \right)
\left[ \frac{(\pa \phi)^3}{x_0-w_0+ix_1} - \frac{(\bar{\pa} \phi)^3}{x_0-w_0
-ix_1} \right] \right. \nonumber \\
&~&~~~~~~~~~~~ \left. + 48 \frac{\pa^2 B}{\pa \phi^2} \left( -\frac{\b^2}{4\pi}
\right)^2 \left[ \frac{(\pa \phi)^2}{(x_0-w_0+ix_1)^2} - \frac{(\bar{\pa}
\phi)^2}{(x_0-w_0-ix_1)^2} \right] \right. \nonumber \\
&~&~~~~~~~~~~~
\left. + 64 \sqrt{2} \frac{\pa^3 B}{\pa \phi^3} \left( -\frac{\b^2}{4\pi}
\right)^3 \left[ \frac{\pa \phi}{(x_0-w_0+ix_1)^3} -
\frac{\bar{\pa} \phi}{(x_0-w_0-ix_1)^3} \right] \right\}
\label{59}
\ena
Using the boundary equation of motion to zero order in
perturbation theory $\left. \frac{\pa \phi}{\pa x_1} \right|_{x_1=0} =0$
and the identity (\ref{47}), one obtains
\EQ
\left. \left\langle (\pa \phi)^4 - (\bar{\pa} \phi)^4 \right\rangle
\right|_{x_1=0}^{(1)} \sim
i \left( 2 \frac{\pa B}{\pa \phi} + 6 \a \frac{\pa^3 B}{\pa \phi^3}
+ 2 \a^2 \frac{\pa^5 B}{\pa \phi^5} \right) (\pa_0 \phi)^3
\label{60}
\EN
Following the same procedure one finds that local contributions
from the second order expansion in the boundary perturbation vanish being
proportional to $\left. \frac{\pa \phi}{\pa x_1} \right|_{x_1=0}$.
The third order contributions give
\bea
&& \left( \frac{1}{\b^2} \right)^3 \int dw_0 dw_0' dw_0''
\left\{ 32 \frac{\pa B}{\pa \phi}(w_0) ~\frac{\pa B}{\pa \phi}(w_0')
\frac{\pa B}{\pa \phi}(w_0'') \left( -\frac{\b^2}{4\pi} \right)^3 \right. \\
&~&~~~~\left. \left( \frac{\pa \phi}{(x-w)(x-w')(x-w'')}
-\frac{\bar{\pa} \phi}{(x-\bar{w})(x-\bar{w}')(x-\bar{w}'')} \right) \right.
\nonumber \\
&& \left. +  96 \frac{\pa^2 B}{\pa \phi^2}(w_0) ~\frac{\pa B}{\pa \phi}(w_0')
\frac{\pa B}{\pa \phi}(w_0'') \left( -\frac{\b^2}{4\pi} \right)^4 \right.
\nonumber \\
&~&~~~~\left. \left( \frac{1}{(x-w)^2(x-w')(x-w'')}
-\frac{1}{(x-\bar{w})^2(x-\bar{w}')(x-\bar{w}'')} \right) \right\}
\nonumber
\ena
and the result is
\EQ
\left. \left\langle (\pa \phi)^4 - (\bar{\pa} \phi)^4 \right\rangle
\right|_{x_1=0}^{(3)} \sim
-2i \left( \left(\frac{\pa B}{\pa \phi} \right)^3 + 6 \a \left(
\frac{\pa B}{\pa \phi} \right)^2 \frac{\pa^3 B}{\pa \phi^3} \right)
\pa_0 \phi
\label{64}
\EN
Contributions from the fourth--order expansion in $B$ give
total $\pa_0$--derivatives.
The same procedure is applied to compute the local contributions
from $\left\langle (\pa^2 \phi)^2 - (\bar{\pa}^2 \phi)^2
\right\rangle$. In this case only the first--order expansion in $B$ gives
non trivial contributions and the bulk equations of motion are used to
evaluate  $\pa^2_1 \phi$ at the boundary.

Finally, adding all the local terms which are not
total $\pa_0$--derivatives, we find the following quantum boundary condition
\bea
&&  \left. \left\langle i(J^{(4)} - \tilde{J}^{(4)} - \Theta^{(4)} +
\tilde{\Theta}^{(4)}) \right\rangle_0 \right|_{x_1=0} = \nonumber \\
&~&~~~~~~~~~~~~~~~~~~~~~~~ \nonumber \\
&~&~~~ = -\frac12
\left[ A \left( \frac{\pa B}{\pa \phi} + 3\a \frac{\pa^3 B}{\pa \phi^3}
+ \a^2 \frac{\pa^5 B}{\pa \phi^5} \right) -2D \frac{\pa^3 B}{\pa
\phi^3} \right] (\pa_0 \phi)^3 \nonumber \\
&~&~~~ + \frac12
\left\{ -\left( D +\a \frac{D}{12} + \a^2 \frac{A}{8} + \a^3 \frac{A}{96}
\right) \frac{\pa^2 V}{\pa \phi^2} \frac{\pa B}{\pa \phi}
+ \left[ 2D -\frac{\a}{12}\left( D+\a^2 \frac{A}{8} \right)  \right]
\frac{\pa V}{\pa \phi} \frac{\pa^2 B}{\pa \phi^2} \right.
\nonumber \\
&~&~~~~~~ \left. + A \left(\frac{\pa B}{\pa \phi} \right)^3 + 3A\a \left(
\frac{\pa B}{\pa \phi} \right)^2 \frac{\pa^3 B}{\pa \phi^3} \right\}
\pa_0 \phi
\label{65}
\ena
In order to cancel the potential anomalies on the right hand side of
this equation, first we need impose the condition
\EQ
A \frac{\pa B}{\pa \phi} + (-2D + 3~\a~A) \frac{\pa^3 B}{\pa \phi^3} +
\a^2~A  \frac{\pa^5 B}{\pa \phi^5} =0
\label{66}
\EN
Using the expressions of the quantum coefficients in (\ref{54}) (cfr. also
(\ref{53}))
we find that the above condition is always satisfied by a boundary
perturbation of the form
\EQ
B = d_0 e^{-\frac{\phi}{2}} + d_1 e^{\frac{\phi}{2}}
\label{67}
\EN
Moreover in this case the term in
(\ref{65}) proportional to $\pa_0 \phi$ is a total $\pa_0$--derivative
and no further restrictions need be imposed on the coefficients $d_j$.
In conclusion for the sinh--Gordon theory the existence of a spin--4 quantum
conserved current does not require quantum corrections to the boundary
perturbation $B$.

\subsection{The $a_2^{(1)}$ model}

For this Toda theory the action in (\ref{1}) is
written in terms of two indipendent
scalar fields.  With a realization of the simple roots
in terms of two dimensional vectors
\EQ
\vec{\a}_1 = (\sqrt{2},0) \qquad \quad \vec{\a}_2 = (-\frac{1}{\sqrt{2}},
-\sqrt{\frac{3}{2}})
\EN
the potential in (\ref{1a}) becomes
\EQ
V = e^{\sqrt{2}\phi_1} + e^{-\frac{1}{\sqrt{2}}\phi_1 -\sqrt{\frac{3}{2}}
\phi_2} + e^{-\frac{1}{\sqrt{2}}\phi_1 +\sqrt{\frac{3}{2}}\phi_2}
\label{68}
\EN
This model has a spin--3 classical conserved current with
the general form
\EQ
J^{(3)} = \frac{1}{3} A_{abc} \pa \phi_a \pa \phi_b \pa \phi_c +
b_{ab} \pa^2 \phi_a \pa \phi_b
\label{69}
\EN
where the coefficients $A_{abc}$ and $b_{ab}$, completely
symmetric and antisymmetric respectively, are determined so that
the classical conservation law (\ref{4}) is satisfied \cite{b2}.
At the classical level, as shown in \cite{b5},
the boundary condition (\ref{9}) fixes the values of
the coefficients, $d_j^2=4$, which appear in the boundary perturbation
$B=\sum_{j=0}^2 d_j e^{\frac12 \vec{\a}_j \cdot \vec{\phi}}$.

Now we study the conservation of the current in (\ref{69})
at the quantum level.
We start considering the conservation law in the upper--half plane and
evaluate $\bar{\pa} \left\langle J^{(3)} \left( -\frac{1}{\b^2} \right)
\int d^2w V \right\rangle$.  Using the massless
propagator (\ref{41}) and following a procedure similar to the one described
for the sinh--Gordon model, we easily find
\bea
&& \bar{\pa} \left\langle J^{(3)} \right\rangle \sim \nonumber \\
&&\sim  \pa \left[ \frac12 b_{ab} V_b \pa \phi_a
+ \frac{\a^2}{48}  A_{abc} \pa V_{abc} \right]
\nonumber \\
&& + \frac12 \left[ A_{abc} V_a + b_{ac} V_{ab} + b_{ab} V_{ac}
+ \frac{\a}{4} A_{abd} V_{acd} + \frac{\a}{4} A_{acd} V_{abd} \right]
\pa \phi_b \pa \phi_c
\label{70}
\ena
where we have defined $V_a \equiv \frac{\pa V}{\pa \phi_a}$ and dropped all
the non--local contributions.
Therefore absence of quantum anomalies in the
conservation of $J^{(3)}$ requires that the terms on the right--hand--side
which are not total $\pa$--derivatives vanish
\EQ
A_{abc}V_a + b_{ac} V_{ab} + b_{ab} V_{ac} + \frac{\a}{4} A_{abd} V_{acd}
+ \frac{\a}{4} A_{acd} V_{abd} =0
\label{71}
\EN
We note that in the classical limit this expression reproduces the
result of \cite{b5}. Introducing the explicit expression of $V$
we compute the quantum coefficients and obtain
\EQ
J^{(3)} = (\pa \phi_1)^2 \pa \phi_2 -\frac{1}{3} (\pa \phi_2)^3
+ \frac{1}{\sqrt{2}}(1+\frac{\a}{2}) \pa^2 \phi_2 \pa \phi_1
- \frac{1}{\sqrt{2}} (1+\frac{\a}{2}) \pa^2 \phi_1 \pa \phi_2
\label{72}
\EN
This current coincides with the one determined in Ref. \cite{b7} up to a total
$\pa$--derivative.
Finally from equation (\ref{70}) we construct the quantum trace
\bea
&& \Theta^{(3)} = -\frac12 b_{ab} V_b \pa \phi_a -
\frac{\a^2}{48} A_{abc} \pa V_{abc} = \nonumber \\
&& =-\frac{1}{2\sqrt{2}} (1+\frac{\a}{2}) V_2 \pa \phi_1
+ \frac{1}{2\sqrt{2}} (1+\frac{\a}{2}) V_1 \pa \phi_2
-\frac{\a^2}{16} \pa V_{112} + \frac{\a^2}{48} \pa V_{222}
\label{73}
\ena
The same procedure can be applied to compute the quantum currents
$\tilde{J}^{(3)}$, $\tilde{\Theta}^{(3)}$ whose expressions are obtained
from (\ref{72}), (\ref{73}) by exchanging holomorphic derivatives with
antiholomorphic ones.

Now we concentrate on the boundary condition (\ref{39}). Thus we consider
\EQ
\left. \left\langle
i(J^{(3)}
-\tilde{J}^{(3)} -\Theta^{(3)} +\tilde{\Theta}^{(3)}) e^{\frac{1}{\b^2}
\int dw_0 B} \right\rangle_0 \right|_{x_1=0}
\EN
where for the time being we leave the coefficients $A_{abc}$ and $b_{ab}$
unspecified. Local corrections to the classical condition (\ref{9})
come from contractions of the currents with the exponential expanded up
to the third order. The calculation is performed along the lines
described in detail for the sinh--Gordon model. Summing
all the contributions the final result is
\bea
&& \left. \left\langle i(J^{(3)}
-\tilde{J}^{(3)} -\Theta^{(3)} +\tilde{\Theta}^{(3)})
\right\rangle \right|_{x_1=0} = \nonumber \\
&& =\frac{1}{\sqrt{2}} \left[
\frac{1}{3} A_{abc} B_a B_b B_c + 2b_{ab} V_a B_b +
\frac{\a^2}{24} A_{abc} V_{abcd} B_d \right] \nonumber \\
&& - \frac{1}{\sqrt{2}} \left[ A_{abc} B_a + 2b_{ab} B_{ac} + 2b_{ac} B_{ab}
+\a A_{abd} B_{acd} + \a A_{acd} B_{abd} \right] \pa_0 \phi_b \pa_0 \phi_c
\label{74}
\ena
In order to cancel the term proportional to $\pa_0 \phi_b \pa_0 \phi_c$
we require
\EQ
A_{abc} B_a + 2b_{ab} B_{ac} + 2b_{ac} B_{ab} +  \a A_{acd} B_{abd}
+ \a A_{abd} B_{acd} =0
\label{75}
\EN
In the classical limit it matches the result in Ref. \cite{b5}.
Comparing (\ref{75}) with (\ref{71}) it is easy to see
that the quantum corrections in both identities
are such that if (\ref{71}) is satisfied with
$V=\sum_{j=0}^2 q_j e^{\vec{\a}_j \cdot \vec{\phi}}$, then (\ref{75})
is also satisfied with
$B = \sum_{j=0}^2 d_j e^{\frac12 \vec{\a}_j \cdot \vec{\phi}}$.
The coefficients $d_j$ are determined and actually acquire an explicit
quantum correction once the other condition
\EQ
\frac{1}{3} A_{abc} B_a B_b B_c + 2b_{ab} V_a B_b + \frac{\a^2}{24}
A_{abc} V_{abcd} B_d =0
\label{76}
\EN
is imposed, with $A_{abc}$ and $b_{ab}$ determined from
eq. (\ref{71}) or equivalently from (\ref{75}).
This relation follows from the requirement of complete
cancellation of local anomalies in (\ref{74}):
it is nonlinear in $B$ and thus it imposes nontrivial constraints on the
coefficients. We note that
setting $\a=0$ the classical result in Ref. \cite{b5} is reproduced, with
$d_j^2 =4$, $j=0,1,2$.
However the presence in (\ref{75}) and (\ref{76}) of quantum corrections
modifies this solution. We solve the equation
(\ref{76}) by introducing the following notation (see Ref. \cite{b5})
\EQ
V = \sum_{j=0}^2 e_j^2 \qquad \quad B = \sum_{j=0}^2 d_j e_j
\label{77}
\EN
where we have defined $e_i = e^{\frac12 \vec{\a}_j  \cdot \vec{\phi}}$.
Moreover we define
\EQ
A_{ijk} \equiv A_{abc} (\a_i)_a (\a_j)_b (\a_k)_c \qquad
b_{ij} \equiv b_{ab} (\a_i)_a (\a_j)_b \qquad C_{ij} \equiv \vec{\a}_i \cdot
\vec{\a}_j
\label{78}
\EN
which, as a consequence of (\ref{71}), satisfy
\EQ
A_{ijk} + b_{ik} C_{ij} + b_{ij} C_{ik} + \frac{\a}{4} [ A_{iij} C_{ik} +
A_{iik} C_{ij} ] = 0
\label {79}
\EN
{}From this equation and the antisymmetry of $b_{ij}$ one
easily derives
\bea
&& A_{iii} =0 \nonumber \\
&& A_{iij} = - A_{jji} = -\frac{2}{1+\frac{\a}{2}} b_{ij}
\label{80}
\ena
for any $i,j = 0,1,2$.
Using the previous relations we rewrite (\ref{76}) as
\EQ
\sum_{i \neq j} \left( - \frac{1}{4} \frac{1}{1+\frac{\a}{2}} d_i^2 +1
\right) b_{ij} d_j  e_i^2e_j =0
\label{81}
\EN
Consequently, in order to maintain at the quantum level the conservation of
the $q^{(2)}$ charge the coefficients $d_j$ must be modified as
\EQ
d_j^2 = 4 \left( 1+\frac{\a}{2} \right) \quad , \qquad j=0,1,2
\label{82}
\EN
This result is exact to all loop orders. Therefore the conservation of the
$q^{(2)}$ charge requires the addition to the action (\ref{1}) of an infinite
number of finite boundary interactions, i.e. a nonperturbative
renormalization of the coefficients $d_j$.

\sect{Conclusions}

We have studied the integrability properties of Toda theories defined in
the upper--half plane presenting a Lax pair approach which allows to
determine the general structure of the boundary perturbation compatible
with the existence of classical higher--spin conserved charges. We have
found that, given the bulk potential
$V= \sum_j q_j e^{\vec{\a}_j \cdot \vec{\phi}}$, the integrable boundary
perturbation is $B=\sum_j d_j e^{\frac{1}{2}\vec{\a}_j \cdot \vec{\phi}}$.
This result generalizes to all Toda models what obtained in Ref. \cite{b5}
for the $a^{(1)}_n$, $d_n^{(1)}$ and $e_6^{(1)}$ systems.
Then we have illustrated the procedure for
the construction of the conserved currents in the interior region supplemented
by the boundary condition.

At the quantum level the conservation laws have been studied using a suitable
generalization of the massless perturbation procedure which is standard
for systems without boundary \cite{b7}.
The requirement of cancellation of local anomalies
leads to a renormalization of the classical currents. We have studied in detail
two explicit examples, the spin--4 current of the sinh--Gordon theory and the
spin--3 current of the $a^{(1)}_2$ Toda model. In the first case quantum
corrections
induce a coupling--constant modification of the current, but no restrictions
need be imposed on the $d_0$, $d_1$ coefficients of the boundary potential.
In the $a^{(1)}_2$ example we have found that in order to insure the quantum
conservation of the corresponding charge, in addition to a renormalization of
the spin--3 current, a finite, nonperturbative renormalization of the
boundary perturbation is necessary. We enphasize that these results are
{\em quantum exact}.
As argued in Ref. \cite{b5} at the classical
level the restriction on the coefficients of the boundary perturbation seems
to be a common feature of all Toda models except for the sinh--Gordon
theory, i.e. $a_1^{(1)}$.
It would be interesting to investigate whether the existence of {\em
quantum} higher--spin conserved charges
requires a finite renormalization of the boundary potential in all
Toda systems.

\end{document}